\documentclass[entropy,article,accept,moreauthors,12pt,a4paper]{mdpi}

\lastpage{x}
\doinum{10.3390/------}
\pubvolume{15}
\pubyear{2013}
\history{Received: 05 September 2013; in revised form: 25 September 2013/ Accepted: 27 September 2013 / \\ Published: xx
}

\usepackage{soul,color}

\Title{The Phase Space Elementary Cell in Classical and \\Generalized Statistics}

\Author{Piero Quarati $^{1,2,}$* and Marcello Lissia $^{2}$}

\address{%
$^{1}$ DISAT, Politecnico di Torino, C.so Duca degli Abruzzi 24, Torino I-10129, Italy\\
$^{2}$ Istituto Nazionale di Fisica Nucleare, Sezione di Cagliari, Monserrato I-09042, Italy; \linebreak E-Mail: marcello.lissia@ca.infn.it}

\corres{E-Mail: pierino.quarati@ca.infn.it; \linebreak Tel.: +39-0706754899; Fax: +39-070510212.}

\abstract{
 In the past, the phase-space elementary cell of a non-quantized system was set equal to the
 third power of the Planck constant; in fact, it is not a necessary assumption. \linebreak We discuss how
 the phase space volume, the number of states and the elementary-cell volume of a system of non-interacting $N$
 particles, changes when an interaction is switched on and the system becomes or evolves
 to a system of correlated non-Boltzmann particles and derives the appropriate expressions.
 Even if we assume that nowadays the volume of the elementary cell is equal to the cube of the Planck constant, $h^3$,
 at least for quantum systems, we show that there is a correspondence between different values of $h$ in the past,
 with important and, in principle, measurable cosmological and astrophysical consequences, and systems with an effective
 smaller (or even larger) phase-space volume described by non-extensive generalized~statistics.
}


\begin{document}

 \vspace{-12pt}
\section{Introduction}

In a microcanonical formulation,
the entropy is proportional to the logarithm of the number of states, $W$, accessible to the systems.

In quantum mechanics, we can count the number of microstates $W$ corresponding to a microstate of total energy, $E$.
This number of microstates $W$ is often called the multiplicity of the microstate.

In classical mechanics, the state of an $N$-particle system is represented
in a $6N$ dimensional \mbox{coordinate-momentum} space ($\Gamma$ space).
Boltzmann solved the problem of infinities when counting the multiplicity of continuous states by
dividing the phase-space in elementary cells, whose volume
$\Delta \Omega$ remains arbitrary in classical mechanics.

The advent of quantum mechanics provided the framework to determine the volume of the smallest elementary cell, which
is given by the cube of the Planck constant: $h^3$.
This lower bound to the volume of the elementary cell is directly connected to the uncertainty principle.

In classical statistical mechanics, the arbitrary volume of the elementary cell does not affect physical quantities.
In fact, as long as the entropy of the system is proportional to the logarithm of the number of states:
$W=\Omega / (\Delta \Omega)^N $
\begin{equation}
 S = k \ln \frac{\Omega}{(\Delta \Omega)^N N!} \quad
\end{equation}
where $\Omega$ is the volume of the phase space accessible to the system, entropy differences:
\begin{equation}
\Delta S = S(A) - S(B)
\end{equation}
do not depend on the volume of the elementary cell $\Delta \Omega$, and therefore, all physical
quantities that depend on entropy differentials, such as the
heat capacity:
\begin{equation}
 C_V = T \left(\frac{\partial S}{\partial T}\right)_{(V,N)}
\end{equation}
or the pressure:
\begin{equation}
P = T \left(\frac{\partial S}{\partial V}\right)_{(E,N)} \quad
\end{equation}
are similarly independent of the choice of the elementary cell volume.

However, the chemical potential:
\begin{equation}
 \mu = - T \left(\frac{\partial S}{\partial N}\right)_{(E,V)}
\end{equation}
does depend on $\Delta\Omega$. Therefore, the tendency of particles to diffuse along the gradient of densities or, more generally, of
chemical potential appears to be affected by the elementary volume of the the phase space.
Furthermore, the elementary cell enters also in the Sackur-Tetrode entropy of a mono-atomic classical ideal gas:
\begin{equation}
 \frac{S}{k N} = \frac{5}{2} + \ln\left[ \frac{V}{N} \left(\frac{E}{N}\right)^{3/2} \frac{ m}{3\pi h^2} \right] \quad
\end{equation}
where $E$ is the internal energy of the gas and $m$ the mass of the particles of the
system. In fact, this entropy incorporates quantum effects through, for instance, the thermal
wave length $ \sqrt{3\pi h^2 N/ (m E)}$.

Up to now, we have considered classical systems composed of non-interacting classical particles or
of particles subjected to short-range forces that do not alter the basic one-particle structure of the
excitations and
that can be absorbed in a few effective parameters.
This kind of systems can be described by the Maxwell-Boltzmann (MB) statistics, and their excitation distribution
is obtained by the maximization~method.

The number of different microstates $w$ accessible to a system of $N$ identical particles is obtained by dividing
by $N!$ the number of microstates $W$ of a corresponding system of distinguishable particles:
\begin{equation}
 w = \frac{W}{N!} \quad
\end{equation}
which is greatly reduced with respect to $W$.
According to Gibbs, all microstates in this microcanonical ensemble have
equal probabilities.

In classical statistical mechanics, the statement $(\Delta \Omega)^N = h^{3N}$
is justified considering the product of an $N$ independent quantum one-particle system
or from the pure dimensional point of view: the Planck constant has dimensions of
the two-dimensional phase-space volume
$x\times p$, and each particle occupies a volume
$\Delta\Omega = h^3$. A possible dimensionless factor cannot be excluded classically.

In conclusion, in classical discrete and continuum statistical mechanics,
the volume of an elementary cell is not determined \textit{a priori}; it is dimensionless in units of
the cube of the Planck constant and could span a large set of values.
The choice that the cell has a volume exactly equal to
$h^3$ is plausible, but only the experiment could possibly determine this constant
in classical mechanics.

There exist complex natural and artificial systems characterized by long-range space-time
correlations whose dynamics cover only a multi-fractal subset of the whole phase space and
develop non-Gaussian correlations, at least for practical finite equilibration times. These
kinds of systems can be effectively described by generalized statistical mechanics that
are a continuous deformation of the usual one~\cite{KaniadakisProc:2002,KaniadakisProc:2004}.
 A particularly simple such generalized
approach is the one that uses the non-extensive Tsallis \mbox{entropy~\cite{Tsallis:1988,TsallisBook:2009}}
 in its continuous and quantum versions.
We have already studied modifications of the phase-space elementary cell in the context of
non-extensive statistical mechanics~\cite{Quarati:2003fp}; this work
develops such an approach on a more general basis.

Recently, S.~Abe \cite{Abe:2010} used this choice $\Delta\Omega=h^3$,
but only for classical discrete systems. He argues that variation of entropy or other measurable quantities
in non-extensive systems do not depend on $h^3$ only in discrete systems. On the contrary,
entropy or other measurable quantities in continuum non-extensive systems depend on $h^3$ against the
hypothesis that the systems are classical. We shall elaborate on this point in Section 3.

In this paper, we show how the phase-space volume can be deformed and how the non-extensive expression
of the number of states changes from the one of an ideal non-interacting particle system. \linebreak
We obtain first the value of the deformed phase space volume.
Then, we deduce the elementary cell volume as a function of the entropic deformation parameter.

Even quantum systems, which appear to have a lower bound to the phase-space volume, due to
the uncertainty principle, could have an effective volume of the elementary cell slightly
smaller than
$h^3$ in non-Boltzmann systems with a fractal measure of the volume. In addition,
there exists the possibility that the Planck constant and, therefore, the volume $h^3$
were different in the past. Both possibilities would have
measurable cosmological and astrophysical consequences.
Entropic uncertainty relations are also~considered.


\section{The Deformation of Phase Space Volume}

The very large number of microstates makes both the phase-space and elementary-cell volume
practically insensitive to small variations of the distribution function. In fact,
the likelihood of deviations is exponentially small, with the exponent proportional to the
large number of degrees of freedom. However interactions can create strong correlations that
strongly reduce the effective number of degrees of freedom relevant at finite time-scales.

Since the equilibrium (maximum) value, $W_0$, is much larger than the values of the
number of microstates $W$ associated with distributions even slightly different from
equilibrium, one usually identifies the state of maximum probability with
the \emph{real state} that is experimentally measured (Sommerfeld \cite{Sommerfeld:book}, p. 218).
However, we will show how going from an extensive classical system to a non-extensive one, \textit{i.e.},
from a pure MB distribution to distributions with power-law tails,
small changes of the phase-space volume can influence the description of the system.

The effect on $\Omega$ of variations from the Maxwellian distribution function $n_{io}$ of
the $i$-th state
can be found in the paper by Bohm and Schutzer~\cite{Bohm:1955}.
In the book of
Diu {\it et al.}~\cite{Diu:1989},
it is shown that in classical physics, the elementary cell is the Planck
constant by comparing the sequence of the admissible states to the sequence of states in quantum mechanics.

Long-range interactions, correlations, memory effects, a fractal space, elementary length or discrete time
are some of the physical reasons systems could not be consistently
described by MB distributions. Such systems can often be better described by
generalized statistics, as developed in the last few years. In~fact, one might also view these effective descriptions as
non-equilibrium metastable states: if the lifetime is long enough, the final thermodynamical equilibrium is not
experimentally relevant.

For the sake of simplicity, we discuss only the so-called $q$-statistics~\cite{Tsallis:1988,TsallisBook:2009}.
 However,
similar conclusions could be reached with other power-law deformations of the MB statistics, such as the
$\kappa$-statistics~\cite{Kaniadakis:2002zz,Kaniadakis:2013},
which offers a mathematically attractive context, symmetry properties,
and interesting relations with the formalism of relativity.

Criticisms to the use of non-extensive, generalized statistics can be found, for instance, in works
by Gross~\cite{Gross:2004}. In Shalizi~\cite{Shalizi:2012}, for instance, one can find criticisms of the
use of approaches that adopt deformed, non-extensive or generalized statistics.

In the following, we show how the phase-space volume varies for small deviations from the standard
distribution function using
the specific formalism of the generalized $q$-statistics.

The phase space volume for a system of $N$ non-interacting Boltzmann particles (or molecules) is defined by:
\begin{equation}
\label{eq:Omega0def}
 \Omega_0 =\frac{N!}{n_{10}! n_{20}! \cdots} \frac{1}{N!} \left( \Delta\Omega_0\right)^N \quad
\end{equation}
for interacting non-Boltzmann particles, we define:
\begin{equation}
\label{eq:OmegaQdef}
 \Omega_q =\frac{N!}{n_{1q}! n_{2q}! \cdots} \frac{1}{N!} \left( \Delta\Omega_q\right)^N \quad
\end{equation}
where $n_{i0}$ and $n_{iq}$ are, respectively, the standard and deformed distributions.
The number of accessible states is:
\begin{equation}
\label{eq:W0Def}
 W_0 =\frac{N!}{n_{10}! n_{20}! \cdots} = N! \frac{\Omega_0}{\left( \Delta\Omega_0\right)^N}
\end{equation}
and:
\begin{equation}
\label{eq:WqDef}
 W_q =\frac{N!}{n_{1q}! n_{2q}! \cdots} = N! \frac{\Omega_q}{\left( \Delta\Omega_q\right)^N}\quad
\end{equation}

We are interested in small deviations from the standard distributions and define:
 $\delta n_{i} = n_{iq} - n_{i0} $.

Using the Stirling formula, $\ln W_q =\ln (N!)-\sum_i \ln (n_{iq}!)$ can be expanded around
the value $W_0(n_{i0})$ that maximizes the phase volume $\Omega_0$ at the MB distribution,
which is expressed as:
\begin{equation}
 n_{i0} = A_M e^{-a_i} \mathrm{\ \ with \ \ } a_i \equiv \beta \epsilon_i \quad
\end{equation}
where
\begin{equation}
 A_M = \frac{N}{V}\left(\frac{\beta}{2\pi m}\right)^{3/2}\Delta\Omega_0
\end{equation}
and the normalization is such that $\sum_i n_{i0} = N $.
In this expansion:
\begin{equation}
 \ln W_q = \left( N\ln N -\sum_i n_{i0} \ln n_{i0}\right) -\sum_i \left( 1+ \ln n_{i0}\right) \delta n_{i}
- \frac{1}{2}\sum_i\frac{\delta n_{i}^2}{n_{i0}} + \cdots
\end{equation}
terms linear in $\delta n_{i}$ vanish, because of the maximum conditions, and retaining the
leading contribution,
the explicit expressions of $W_q$ and $\Omega_q$ for a gas that deviates only slightly from the MB distribution,
assuming for the moment an unchanged elementary cell,
 are:
\begin{equation}
\label{eq:Wq}
 W_q = W_0 \exp\left(- \frac{1}{2}\sum_i\frac{\delta n_{i}^2}{n_{i0}} \right) \quad
\end{equation}
and:
\begin{equation}
 \Omega_q = \Omega_0 \exp\left(- \frac{1}{2}\sum_i\frac{\delta n_{i}^2}{n_{i0}} \right) \quad
\end{equation}

The deformed distribution is:
\begin{equation}
 n_{iq} = A_M A_q\; e^{-\beta \epsilon -(1/2)(1-q) (\beta \epsilon )^2 } \quad
\end{equation}
with $\sum_i n_{iq} = N$ where:
\begin{equation}
 A_q = 1 + \frac{15}{8}(1-q) + o(1-q)^2 \quad
\end{equation}

For $q\approx 1$, we have:
\begin{equation}
 n_{iq} = n_{i0} \left( 1+ \frac{15}{8}(1-q) - \frac{(1-q)}{2} (\beta\epsilon_i)^2\right)
\end{equation}

\begin{eqnarray}
\nonumber \sum_i \frac{(\delta n_{i})^2}{n_{i0}} &=& \frac{(1-q)^2}{4} \sum_i n_{i0} \left[\frac{15}{4} - (\beta \epsilon_i)^2\right]^2 \\
 &=& N \frac{(1-q)^2}{4} \left[ \left(\frac{15}{4}\right)^2 -\frac{15}{2} \langle (\beta \epsilon_i)^2 \rangle
 + \langle (\beta \epsilon_i)^4 \rangle \right] \\
\nonumber &=&
 N \frac{(1-q)^2}{4} \left[ \left(\frac{15}{4}\right)^2 -\frac{15}{2} \frac{15}{4}
 + \frac{945}{16} \right] = \frac{45}{4} N (1-q)^2 \quad
\end{eqnarray}
where
in the preceding equations, we have calculated the average values with the \linebreak unperturbed distribution.

Therefore, we obtain:
\begin{equation}
\label{eq:WqDeform}
 W_q = W_0 e^{-(45/8) N (1-q)^2} \quad
\end{equation}
From $W_q/W_0 = \left( \Omega_q / \Omega_0 \right) \left(\Delta\Omega_0 / \Delta\Omega_q\right)^N$ (see, for instance, Equations
(\ref{eq:W0Def}) and (\ref{eq:WqDef})), this change can be interpreted microscopically as a change of an elementary cell
leaving the phase-space \linebreak volume unchanged:
\begin{equation}
\label{eq:DeltaOmegaQnoQpower}
\Delta \Omega_q = \Delta\Omega_0 e^{(45/8) (1-q)^2} \quad
\end{equation}
Because the correction factor to $W_q$ depends on $(q-1 )^2$, $\Delta\Omega_q$ is always larger
than $\Delta\Omega_0$.
However, if we had assumed that the number of accessible states were independent of $q$, $\Omega_q$ and
$(\Delta\Omega_q)^N$ would change in the same way, and one could obtain an elementary cell smaller than
$\Delta\Omega_0$; in particular, \linebreak $\Delta \Omega_q = \Delta\Omega_0 \exp(-(45/8) (1-q)^2$.
Here, we just remark that $W_0$ and $W_q$ have different values
only to the second order in an expansion in $(1-q)$.
In the definition of the standard entropy, we use $W_0$, while in the definition of the
$q$-entropy, we should use the $q$-logarithm of $W_q$. However, the leading linear correction comes only from the $q$-logarithm,
and one can use $W_0$.


\section{Classical Volume Cell, Continuous States and Non-Maxwellian Statistics}

When computing the entropy of a classical system with states parameterized by continuous parameters,
coarse graining, \textit{i.e.} a finite cell size, overcomes the difficulties of
defining the correct measure of probability densities.
Abe~\cite{Abe:2010} argues that generalized statistical mechanics with non-logarithmic entropies
can be applied only to physical systems with discrete degrees of freedom.
In fact, the elementary volume necessary to regularize the sum over states in
continuous Hamiltonian systems does not disappear in the final results for non-additive entropic functions:
if the elementary cell is assumed to be $h^3$ from the underling quantized formulation, classical results
would contain the quantum constant $h^3$.
This reasoning spurred an interesting discussion~\cite{Andresen:2010,Abe:2010c} between Abe and Andresen.

We remark that a discrete cell can appear also in classical mechanics, for instance, due to effective
coarse graining with a volume not necessarily equal to $h^3$.
In addition, the same constant $h$ can have an interpretation also in a classical
context~\cite{Boyer:1969zza,Boyer:2008bi,Boyer:2012}.

We conclude that both continuous and discrete classical systems can contain a phase space
elementary cell that in classical physics can have any value: in particular, its volume
can be $h^3$.
Therefore, both continuous and discrete system can be described by generalized statistical
mechanics that permit elementary cells with a volume different from $h^3$.


\section{The Generalized Number of States and the Generalized Elementary Cell Volume}

We consider the standard classical phase space $\Omega_0$, which is maximum when the global \linebreak
thermodynamical equilibrium distribution $n_{i0}$ is the MB distribution
for the $i$-th state. The elementary phase space cell is $\Delta\Omega_0$, and $N$ is the
number of particles of the system.

Let us define the dimensionless quantity:
\begin{equation}
\label{eq:MqOmega0delta0}
 M_0^N \equiv \frac{\Omega_0}{(\Delta\Omega_0)^N } = \frac{W_0}{N!} =
 w_0
\end{equation}
which is the measure of the $3D$ phase-space volume relative to the volume of the one-particle
cell $\Delta\Omega_0$ and where $w_0$ is the number of microstates.

Analogously in the $q$-deformed formalism, we define the $q$-volume $\Omega_q$ and the $q$ elementary cell
$\Delta\Omega_q$ and introduce the dimensionless measure:
\begin{equation}
\label{eq:MqOmegaQdeltaQ}
 M_q^{\otimes_q N} = \frac{\Omega_q}{(\Delta\Omega_q)^N } =
\frac{W_q}{N!} =
 w_q
\quad
\end{equation}
where we have used the $q$-power (see, for instance, \cite{Chung:2013}):
\begin{equation}
 M_q^{\otimes_q N} = \left[ N M_q^{(1-q) }-N+1\right]^{1/(1-q)} \quad
\end{equation}

The expansion of $ M_q^{\otimes_q N} $ for $N\gg1$ around the value $q = 1$ (see Appendix~\ref{app:expansion} for details)
gives:
\begin{eqnarray}
\label{eq:MqNdef}
 M_q^{\otimes_q N} &=&
 M_q^N
 \left[ 1 + \frac{q-1}{2}N^2 \left( \ln M_q \right)^2
 \right]
 \\
 &=&
 \frac{W_q}{N!} \times\left[1+\frac{q-1}{2}
 \left( \ln\frac{W_0}{N!}\right)^2 \right]
\quad
\end{eqnarray}
where, inside the logarithm, we have used $M_q=M_0$ to the order in $1-q$ considered.
If we divide the above equation by the corresponding equation with the index of zero
($q=1$), we find:
\begin{equation}
 \frac{M_q^{\otimes_q N}}{M_0^N} =
 \frac{W_q}{W_0} \times\left[1+\frac{q-1}{2}
 \left( \ln\frac{W_0}{N!}\right)^2 \right] = \left(\frac{\Omega_q}{\Omega_0}\right) \left( \frac{\Delta\Omega_0}{\Delta\Omega_q} \right)^N
\quad
\end{equation}
where the last equality derives from definitions (\ref{eq:MqOmega0delta0}) and (\ref{eq:MqOmegaQdeltaQ})
of $M_0^N$ and $M_q^{\otimes_q N}$. Therefore:
\begin{equation}
\Delta\Omega_q = \Delta\Omega_0 \left( \frac{W_q}{W_0} \right)^{-1/N} \left[1+\frac{q-1}{2}
 \left( \ln\frac{W_0}{N!}\right)^2 \right]^{-1/N} \left(\frac{\Omega_q}{\Omega_0}\right)^{1/N}
\quad
\end{equation}

Inserting the deformed expression of $W_q$, as derived in Equation~(\ref{eq:WqDeform}),
and ascribing the whole dependence on $q$ to the micro-physics, \textit{i.e.} $\Omega_q=\Omega_0$,
as it appears reasonable if the thermodynamical limit exists,
 we obtain, finally:
\begin{equation}
 \Delta\Omega_q = \Delta\Omega_0 \; e^{(45/8)(q-1)^2} \times
 \left[1+ \frac{q-1}{2} \left(\ln\frac{W_0}{N!}\right)^2 \right]^{-1/N} \quad.
\end{equation}
The expression of the $q$-deformed elementary cell is made of two factors:
the first is the elementary cell derived without the use of $q$-algebra (see Equation~(\ref{eq:DeltaOmegaQnoQpower}));
the second represents the correction due to the $q$-algebra.
Note that the last relation shows how $\Delta\Omega_q$ can be smaller or larger than $\Delta\Omega_0$,
depending on the value of $q$.

In Appendix~\ref{app:numerical}, we also consider the corresponding entropy with
a few numerical examples.

\section{Entropic Uncertainty Relations and Elementary Cell}

The above sections show the link between elementary-cell
volume and the uncertainty principle.
It is well known that generalizations and extensions
of the original uncertainty relations of Heisenberg
have been proposed involving Shannon, R\'enyi and, later,
Tsallis entropy.
To be more clear: quantum mechanical uncertainty relations for momentum and position
were expressed as inequalities in terms of Shannon, R\'enyi or Tsallis entropy.

Uncertainty phase space volume is related to the phase-space elementary cell
that therefore can be expressed in terms of Shannon, R\'enyi and Tsallis entropy, giving us
the elementary volume for these generalized
statistics~\cite{Partovi:1983,BialynickiBirula:1984,Rajakopal:1995,BialynickiBirula:2006,%
Wilk:2009,Wilk:2011,BialynickiBirula:2011a,BialynickiBirula:2011b}.

Let us introduce the experimental phase-space elementary cell volume
$\Delta\Omega_E$ as the volume of phase space determined by the resolution
of the measuring instruments. On the other hand, the elementary cell volume,
as discussed in the previous sections, is related to the Heisenberg uncertainty principle
and does not depend on the accuracy of our measuring instruments.

In the Shannon information formalism, the entropy is:
\begin{equation}
 H=-\sum_k p_k \ln p_k
\end{equation}
where $p_k$ is the dimensionless position distribution for the $k$ state ($\kappa_B=1$).
The quantity, $H$, is the position for Shannon. Analogously, $H$ can also be expressed for momentum.

In the R\'enyi framework \cite{BialynickiBirula:2006}, position and momentum entropy have the expressions:
\begin{eqnarray}
H_{\alpha}^x=\frac{1}{\alpha}\ln\sum_k p_k^{\alpha} \quad\\
 H_{\beta}^p=\frac{1}{\beta}\ln\sum_k p_k^{\beta} \quad
\end{eqnarray}
where $H_{\alpha}^x$ and $H_{\beta}^p$ represent the uncertainty in position and
momentum for R\'enyi statistics in the quantum world. When
the entropic parameter $\alpha$ or $\beta$ goes to one, one recovers the Shannon entropy.

Shannon connected the content of the measure of information with a probability distribution, inserting
the set of probabilities obtained from quantum mechanics into the information entropy.

We define the dimensionless relative volume of the cell:
$\omega_S$
\begin{equation}
 \omega_S=\frac{\Delta\Omega_E}{\Delta\Omega_0} =\left(\frac{\delta x \delta p}{h} \right)^3
\end{equation}
This quantity decreases as we enter into the quantum regime.
In the Shannon frame, we have identified $h^3$ with the standard elementary cell
$\Delta\Omega_0$.

From the entropic uncertainty relations, we can give analytical expressions of
$\omega$ in the Shannon, R\'enyi and Tsallis statistics.

In quantum mechanics and in information theory, the entropic uncertainty
is defined as the sum of temporal and spectral entropies.
The uncertainty principle of Heisenberg can be expressed as a lower bound on the sum of these
entropies. This is much stronger than the standard statement of the uncertainty Heisenberg
principle in terms of the product of standard deviations.

The Shannon entropic uncertainty relation can be written:
as~\cite{BialynickiBirula:2006}
\begin{equation}
H^x + H^p \geq 1 - \ln2 - \frac{1}{3}\ln \omega_S \quad
\end{equation}
By inverting the above relation, we obtain:
\begin{equation}
 \omega_S \geq e^{3-\ln8} e^{-3(H^x + H^p)} \quad
\end{equation}

The lower quantum mechanics bound on $(H^x + H^p)$ is found for
$\omega = 1$: $(H^x + H^p) \geq 1-\ln2\approx 0.307$.
Corresponding to $\omega <1 (>1)$, one has lower bounds greater (smaller) than
$1-\ln2$.

Let us report the following limiting cases
(these limits are unphysical, because they violate
the Heisenberg principle):
\begin{itemize}
 \item if $\Delta\Omega_E$ goes to zero,
 then $( H^x + H^p )$ goes to infinity (complete lack of information);
 \item if $( H^x + H^p )=0$ (perfect information), then $\omega_S = \exp(3 - \ln 8 )$.
\end{itemize}

The R\'enyi entropic relation is:
\begin{equation}
H^x_\alpha + H^p_\beta \geq
-\frac{1}{2}\left( \frac{\ln\alpha}{1-\alpha} + \frac{\ln\beta}{1-\beta} \right)
-\ln2-\frac{1}{3}\ln\omega_R \quad
\end{equation}
where $\omega_R = \Delta\Omega_E / \Delta\Omega_R$,
$\Delta\Omega_R$ is the elementary cell in the R\'enyi frame,
and the entropic parameters must verify $\alpha^{-1}+\beta^{-1}=2$.
Therefore, from Equation~(7) of~\cite{BialynickiBirula:2006}, we derive:
\begin{equation}
 \omega_R \geq \frac{1}{8} e^{-3(H^x_\alpha + H^p_\beta)}
 e^{-\frac{3}{2}\left( \frac{\ln\alpha}{1-\alpha} + \frac{\ln\beta}{1-\beta} \right)}
\end{equation}

If the same measure of uncertainty is used for both variables,
 position and momentum, we obtain from Equation~(46) of~\cite{BialynickiBirula:2006}:
\begin{equation}
 \omega_{\hat{R}} \geq \frac{1}{8} e^{-3(H^x_\beta + H^p_\beta)} \frac{1}{\beta^3} \left(2\beta-1 \right)^{(3/2-3\beta)/(1-\beta)}
\end{equation}
with $1/2\leq\beta\leq 1$.

The cell $\Delta\Omega_R$
 can easily be derived in terms of the sum of position
and momentum uncertainty and of the experimental phase-space elementary cell volume.

A detailed analysis in terms of the phase-space elementary cell needs to be performed.
At the moment, we say that the volume of the elementary cell depends on the effective
statistics relevant for the specific physical system, and its value in units of $h^3$
can be smaller than one; in some cases, this effective volume can be even much bigger than
one. In addition, the Planck constant, and, therefore, the fine-structure constant
$\hbar c / e^2\approx 137.036$, could have had values different from the one
presently measured.

Existing generalizations of the uncertainty principle do not make reference to the
implications for the volume of the elementary cell
in quantum dynamics. Parallel modifications of the elementary cell can be
studied within classical dynamics, where the scale $h^3$ is not derived from an
uncertainty principle. \linebreak For instance T.~H.~Boyer interprets
$h$ as the scale of random classical zero-point radiation that is one of the
solutions of the Maxwell equations \cite{Boyer:1969zza,Boyer:2008bi,Boyer:2012}.

We stress the important relation between non-extensive statistics and modifications of
the elementary volume of phase space.

In the following, we summarize uncertainty entropic relations derived from
Bialynicki-Birula \cite{BialynickiBirula:2006}
and Wilk \cite{Wilk:2009,Wilk:2011}
using also the condition of the non-negativity of momentum entropy and position entropy
and of the right-hand sides of inequalities {(7) and (46)} of \cite{BialynickiBirula:2006}
and (19) and (20) of~\cite{Wilk:2009}

For Shannon, we have:
\begin{equation}
\frac{2}{e} e^{(H^x+H^p)} \geq \frac{\Delta\Omega_0^{1/3}}{\delta x \delta p} \geq \frac{2}{e}
\end{equation}

For R\'enyi, we have:\\
from Equation (7) of \cite{BialynickiBirula:2006} if $\alpha \neq\beta$ with $\alpha \geq 1$ and $\beta \leq 1$:
\begin{equation}
2e^{\frac{1}{2}\left(\frac{\ln\alpha}{1-\alpha}+\frac{\ln\beta}{1-\beta}\right)}
 e^{(H^x_\alpha+H^p_\beta)}
\geq \frac{\Delta\Omega_R^{1/3}}{\delta x \delta p}
\geq 2e^{\frac{1}{2}\left(\frac{\ln\alpha}{1-\alpha}+\frac{\ln\beta}{1-\beta}\right)}
\end{equation}
where $1/\alpha + 1/\beta = 2$;\\
from Equation (46) of \cite{BialynickiBirula:2006}, if $\alpha=\beta$:
\begin{equation}
2\beta(2\beta-1)^{\frac{\beta-1/2}{1-\beta}}
 e^{H^x_\beta+H^p_\beta}
\geq
\frac{\Delta\Omega_{\hat{R}}^{1/3}}{\delta x \delta p}
\geq
2\beta(2\beta-1)^{\frac{\beta-1/2}{1-\beta}}
\end{equation}
where $1/2 \leq\beta \leq 1$.

For Tsallis, we have two relations with $\alpha \neq\beta$ ($\alpha > \beta$):
if $\eta < 1$ (from Equation (19) of~\cite{Wilk:2009}):
\begin{equation}
 2\beta \left(\frac{\beta}{\alpha}\right)^{\frac{1}{2(\alpha-1)}}
\geq
\frac{\Delta\Omega_q^{1/3}}{\delta x \delta p}
\geq
 2\beta \left(\frac{\beta}{\alpha}\right)^{\frac{1}{2(\alpha-1)}}
\left[
 \frac{1}{1+(1-\alpha)(H^x_\beta+H^p_\beta)}
\right]^{\alpha/(\alpha-1)}
 \quad
\end{equation}
while if $\eta > 1$ (from Equation (20) of \cite{Wilk:2009}):
\begin{equation}
 2\beta \left(\frac{\alpha}{\beta}\right)^{\frac{1}{2(1-\alpha)}}
\geq
\frac{\Delta\Omega_q^{1/3}}{\delta x \delta p}
\geq
 2\beta \left(\frac{\alpha}{\beta}\right)^{\frac{1}{2(1-\alpha)}}
 \left[ \frac{1}{1+(\alpha-1)(H^x_\beta+H^p_\beta)}\right]^{\alpha/(\alpha-1)}
\quad
\end{equation}
where:
\begin{equation}
\eta(\alpha,\beta) = \left(\frac{\beta}{\alpha}\right)^{1/(2\alpha)}
\left(2\beta\right)^{(\alpha-1)/\alpha}
\left(\frac{\delta x \delta p}{\Delta\Omega_q^{1/3}}\right)^{(\alpha-1)/\alpha}
\quad
\end{equation}

The distribution of momentum and position with different entropic parameters
are taken with the same elementary cell $\Delta\Omega_q$.

The meaning of the above relations can be summarized as follows.
Once the effective statistical framework (Shannon, R\'enyi, Tsallis or others),
and, therefore, the corresponding volume of the elementary cell, is determined by the physical characteristics of the system,
the only meaningful measurements of two conjugate quantities, such as
position and momentum, are those whose uncertainty product satisfies the above relations.

\section{Astrophysical and Cosmological Implications}

Experiments have been arranged in astrophysics to measure the cosmic red shift, the variation of the
fine-structure constant in space and time, and to observe a cosmological rate of change of the Planck
constant: results are under scrutiny and do not appear to give conclusive answers.
Cosmic acceleration should be verified by future ground-based laboratory experiments, where
dark matter and dark energy are looked~for.

It is believed that phase-space structures with a volume smaller than
$h^3$ do not exist or do not matter; however, Zurek \cite{Zurek:2001}
has shown that sub-Planck structures exist in the quantum version of classically chaotic systems.
This situation is allowed in non-extensive statistical mechanics~\cite{Wolf:2002},
 where an elementary cell
smaller than the Planck constant can be admitted at particular values of the entropic parameters.

A conjecture on a hierarchical cosmological time evolution of the Planck constant has been accepted.
This hypothetical phenomenon might be responsible for the cosmological high redshifts as an alternative
mechanism to the Doppler effect. The Bohr radius of a hydrogen atom and the probability of
the alpha decay of atomic nuclei have been estimated: both results seem to be compatible
with observational data concerning the light elements' abundance in the early Universe \cite{Avetissian:2009}.

Several laboratory measurements of the variation in time of the fine-structure constant
assume that the variation is entirely caused by the dependence on time of the Planck constant.
Laboratory measurements are reviewed by J. P. Uzan \cite{Uzan:2003}.

Insofar as the microscopic world is described by the principles of quantum mechanics, physical
results must follow the
uncertainty principle of space coordinate and momentum and depends on the Planck constant:
for instance, atomic and nuclear energy levels are functions of the
fine structure constant.

On the other side, the uncertainty principle and the Planck constant are not confined to microscopic
physics, but they also determine the fundamental characteristics of the
macroscopic physics, including the astrophysical phenomena of large objects, cosmic radiation or
the cosmological evolution of the~universe.

Laboratory experiments measure the present value of the Planck constant $h$ and,
indirectly, the corresponding quantum phase-space elementary cell. However, there exist
recent astrophysical observations that should test
the value of $h$ in the past~\cite{Uzan:2003,Webb:2010hc,Seshavatharam:2013}.
These first observations suggest that the Planck constant was smaller in the past.

If this smaller Planck constant must apply also to the elementary cell of the phase space, we could describe
these results by a generalized statistical mechanics caused, for instance, by non-ideal conditions in the past.

Therefore, we may justify an elementary cell or a Planck constant different (smaller) than the actual~value:
\begin{itemize}
 \item by
observing the cosmological rate of change of the Planck constant, future cosmic acceleration can be verified
from ground-based laboratory experiments~\cite{Seshavatharam:2013};
\item by observations, suggesting a smaller value of the fine structure constant at high redshift~\cite{Webb:2010hc};
\item by the classical interpretation~\cite{Boyer:1969zza,Boyer:2008bi,Boyer:2012} of the constant $h$;
\item by a fractal space-time~\cite{Pellicer:2006}.
\end{itemize}

\section{Conclusions}

In this work, we have shown that for classical systems non-describable by the Maxwell-Boltzmann-Gibbs statistics,
but by generalized statistics (like the Tsallis $q$-statistics),
the number of accessible states
and the elementary cell volume differ from the ideal case:
the deformed phase-space elementary volume is always larger than the ideal one
independently of the sign of $1-q$,
because the leading correction depends on $(1- q )^2$,
while the number of states and the elementary cell are smaller or
larger, depending on $q$.
In classical systems, the elementary cell is not a universal constant, and the elementary cell
can have a volume slightly smaller than $h^3$ in spite of the uncertainty principle.

If we assume that the volume of the elementary cell is $h^3$, this cell could have been
smaller in the past because $h$ was smaller for its intrinsic evolution or, \textit{vice versa},
one could correlate the evolution of $h$ to the change of the effective volume of the
elementary cell, due to stronger spatial and temporal correlations or the fractal measure
of space when densities and interactions were stronger.
These interpretations are especially
interesting if the need for smaller values of the Planck
constant in the past to explain some astrophysical effects,
de-coherence effects and electromagnetic problems is confirmed.

We have shown results only in the limit of small deviations from the standard dynamics, since they
seem to be the most relevant for possible applications, but the qualitative considerations
and results would be similar without these simplifying approximations.

\section*{\noindent Conflicts of Interest}
\vspace{12pt}

The authors declare no conflict of interest.

\bibliographystyle{mdpi}
\makeatletter
\renewcommand\@biblabel[1]{#1. }
\makeatother

\newpage

\appendix
\section{Appendix}
\label{app:expansion}

In this Appendix, we give a few more details about the expansion of
$M_q^{\otimes_q N}$ for values of $q$ close to one.

From the definition of the $q$-power \cite{Chung:2013}:
\begin{equation}
\label{eq:defNqPowerM}
 M^{\otimes_q N} = \left[ N M^{(1-q) } -N+1 \right]^{1/(1-q)}
\end{equation}
we expand the
logarithm of the $N$-th $q$-power of $M$
up to the leading correction in $x\equiv (1-q)\ln M$:
\begin{eqnarray}
\ln\left( M^{\otimes_q N} \right)
 &=&
 \frac{\ln M}{x} \ln \left[ 1 + N \left( e^x -1 \right) \right]
 = \frac{\ln M}{x} \left[ N \left( x + \frac{x^2}{2} \right) - \frac{(Nx)^2}{2} + o(x^3) \right] \\
 &=&
\label{eq:expanionPowerQ}
\ln M^N - \frac{x(N-1)}{2} \ln M^N + o(x^2) \quad
\end{eqnarray}
Considering the case $N\gg 1$, taking the exponential and substituting back
$x= (1-q)\ln M$, the result~is:
\begin{equation}
 M^{\otimes_q N} = M^{N} \left[ 1 - \frac{x N^2}{2}\ln M + o(x^2) \right] =
 M^{N} \left[ 1 + \frac{(q-1)}{2} (\ln M^N)^2+ o((1-q)^2) \right] \quad
\end{equation}
Note that in the opposite limit $N=1$, the first correction in $q-1$ vanishes (see Equation (\ref{eq:expanionPowerQ}));
in fact, the case $N=1$ does not depend on $q$:
\begin{equation}
 M^{\otimes_q 1} = \left[ M^{(1-q) }\right]^{1/(1-q)} = M \quad
\end{equation}

Since the difference between $M_0$ and $M_q$ is of the order $(1-q)^2$ (see Equation~(\ref{eq:WqDeform})
and consider that $W=N! M^N$), we can take
$M^N=M_q^N=M_0^N= \Omega_0 / (\Delta\Omega_0)^N$ inside the logarithm and consider only terms up to $1-q$:
\begin{equation}
 M^{\otimes_q N} =
 \frac{\Omega_q}{(\Delta\Omega_q)^N} =
\frac{\Omega_0}{(\Delta\Omega_0)^N}
\left[ 1 + \frac{q-1}{2} \left(\ln\left(\frac{\Omega_0}{(\Delta\Omega_0)^N}\right)\right)^2\right]
\end{equation}

The expression of the number of available states becomes:
\begin{equation}
 W_q = N! \frac{\Omega_q}{(\Delta\Omega_q)^N}
 = N! \frac{\Omega_0}{(\Delta\Omega_0)^N}
 \left[ 1+ \frac{q-1}{2} \left(\ln\left(\frac{\Omega_0}{(\Delta\Omega_0)^N}\right)\right)^2\right]
\end{equation}

or, in other notation:
\begin{equation}
 W_q = W_0 \left[ 1+ \frac{q-1}{2} N^2 \left(\ln\left(\frac{W_0}{N!}\right)^{1/N}\right)^2\right]
 = W_0 \left[ 1+ \frac{q-1}{2} \left(\ln\left(\frac{W_0}{N!}\right) \right)^2\right] \quad
\end{equation}

\section{Appendix}
\label{app:numerical}

In this Appendix, as an application of the expressions
reported in the main text based on the \linebreak
q-definition of the elementary cell,
we give the expression of the entropy $S_q$ using the rules of
$q$-algebra for $ M_q^{\otimes_q N}$:
 \begin{equation}
 \frac{S_q^{(1)}}{k} =\ln_q w_q \approx \frac{\left(M_q^{\otimes_q N}\right)^{1-q}-1}{1-q} \quad
 \end{equation}

In the case of one particle:
\begin{equation}
 \frac{S_q^{(1)}}{k} = \frac{W_0 ^{1-q}-1}{1-q}
\end{equation}
Using as suggested by Sommerfeld~\cite{Sommerfeld:book} for one particle $W_0=3\cdot 10^{4}$,
let us give as examples:
\begin{eqnarray}
\frac{1}{k} S^{(1)}_{q=1.02} &=& 9.32 \\
\frac{1}{k} S^{(1)}_{q=1.005} &=& 10.05 \\
\frac{1}{k} S^{(1)}_{q=1} &=& \ln W_0 =10.31 \quad
\end{eqnarray}
we see that $S_q$ decreases as $q$ increases from $q=1$. In fact, for $|1-q|\ll 1 / \ln W_0 \approx 0.1$, one
finds $S^{(1)}_q / k \sim \ln W_0 \left[ 1 - (1/2)(q-1)\ln W_0 \right]$.

We calculate now the $q$-entropy for $N$ identical particles using the $q$-algebra
for the sum:
\begin{equation}
 \frac{1}{k} S_q(N) = \frac{1}{k} S_q^{\oplus_N}(1) = \frac{1}{1-q}
\left\{\left[W_0 \left(1+\frac{q-1}{2}\ln W_0 \right) \right]^{(1-q)N} -1\right\} \quad
\end{equation}
For large $N$ and $q>1$, or, better, for $N (q-1) \gg 1 / \ln W_0$, which means in our case
$N\gg 1 / (10(q-1)$, one finds $S_q(N) / k \approx 1/(q-1)(1-\exp(-(q-1)N\ln W_0)) $. In the opposite limit,
$N |q-1| \ll 1 / \ln W_0$, which means in our case
$(q-1)\ll 1 / (10 N)$, one finds $S_q(N) / k \approx N \ln W_0 -(q-1)N\ln A (N\ln A -1) /2$. In general, the entropy
grows linearly as $N\ln W_0$, as long as it is smaller than $1/(q-1)$; then, it saturates to the
constant $1/(q-1)$. As numerical examples for the chosen $W_0=3\cdot 10^4$:
\begin{eqnarray}
 q=1.02 \quad \quad \quad \frac{1}{k} S_q(N) &\approx& \frac{1}{q-1} = 50 \\
 q=1.005 \quad \quad \quad \frac{1}{k} S_q(N) &\approx& \frac{1}{q-1} = 200 \\
 q=1 \quad \quad \quad \frac{1}{k} S_q(N) &=& N\ln W_0 \approx 10.31 N \quad
\end{eqnarray}


\end{document}